\documentclass[aps,prc,superscriptaddress,showpacs,showkeys,a4paper,twocolumn]{revtex4}
\usepackage{graphicx}
\begin{document}
\title{Extended Longitudinal Scaling and the Thermal Model}
\author{J.~Cleymans}
\email{Jean.Cleymans@uct.ac.za}
\author{J.~Str\"umpfer}
\email{johans@mighty.co.za}
\affiliation{UCT-CERN Research Centre and Department  of  Physics,\\
 University of Cape Town, Rondebosch 7701, South Africa}
\author{L.~Turko}
\email{turko@ift.uni.wroc.pl}
\affiliation{UCT-CERN Research Centre and Department  of  Physics,\\
 University of Cape Town, Rondebosch 7701, South Africa}
\affiliation{Institute of Theoretical Physics, University of Wroclaw,\\
Pl.~Maksa Borna 9, 50-204  Wroc{\l}aw, Poland}
\date{\today}
\begin{abstract}
The property of  extended longitudinal scaling of rapidity distributions was noticed
recently over a broad  range of beam energies.
It is shown here that this property is consistent with predictions of the
statistical thermal model up to the highest RHIC beam energies, however, we expect that at LHC energies
the rapidity distribution of produced particles will violate extended longitudinal scaling.
\end{abstract}
\maketitle
%
%
%
%
\section{Introduction}
It was widely expected \cite{Bjorken:1982qr} that the rapidity distribution of particles
produced in relativistic heavy-ion collisions would show  a plateau around central
rapidities. Although it was not observed at SPS energies, there was a rather general belief
about its existence in RHIC experiments. Now, as the final countdown \cite{Abreu:2007kv} to LHC has started,
expectations are much more cautious \cite{plateau}. Recent results about the rapidity of charged mesons \cite{Bearden:2004yx}
and pseudorapidity  distributions \cite{Back:2004je, Alver:2007we} do not allow for any firm prediction
concerning the existence of a plateau at LHC energies.
Instead of this, a new property emerging -  extended longitudinal scaling in rapidity distributions.
The shape of the pseudorapidity distribution scales according to the limiting fragmentation hypothesis.
The distributions of particle yields are largely independent of energy over a broad region of
rapidity when viewed in the rest frame of one of the colliding particles. In this kinematic region
it is allowed to neglect differences between pseudorapidity and rapidity distribution.

Extended longitudinal scaling was observed  in high energy $pp\ $ collisions \cite{Alner:1986xu} and is
also a property of ultrarelativistic heavy-ion collisions.
In this paper we show that the
extended longitudinal scaling feature of the shifted rapidity distribution  also arises within the thermal model
up to the highest RHIC energies. However, when an extrapolation is made to LHC energies the
extended longitudinal scaling effect vanishes. This would violate some of LHC predictions based on the
extended longitudinal scaling feature \cite{Busza:2004mc}.

\section{Rapidity distributions}

The statistical thermal model has been recently extended \cite{Biedron:2006vf, becattini} to allow
for the description of the rapidity distribution of produced particles in heavy-ion collisions.
Chemical potentials and the temperature become rapidity dependent quantities. This property corresponds
to the changing nature of the expanding fireball.

An extension to the thermal model \cite{becattini} is used to calculate the rapidity distributions.
The model uses a Gaussian distribution of fireballs  centered at zero and described by
\begin{equation}
\rho\left(y_{FB}\right)=\frac{1}{\sqrt{2 \pi}\sigma} \exp\left( -\frac{y_{FB}^2}{2\sigma^2}\right).
\end{equation}
The rapidity distribution of particle $i$ is then calculated by
\begin{equation}
\frac{d N^i}{dy} =
\int\limits_{-\infty}^{+\infty}\rho\left(y_{FB}\right)\frac{d N_1^i\left(y-y_{FB}\right)}{dy} dy_{FB}.
\end{equation}
where $\frac{d N_1^i}{dy}$ is obtained from the thermal distribution of hadrons from a single fireball.
It is necessary to assume universality of the chemical freeze-out conditions.
This means that the temperature and the baryonic chemical potential are
related via the freeze-out curve deduced from particle yields at varying beam energies.
A parametrization of the universal freeze-out curve is given by \cite{clorw}
\begin{equation}\label{freeze}
    T = 0.166 - 0.139\mu_B^2 - 0.053\mu_B^4  .
\end{equation}

The extension of the thermal model introduces a new energy dependent parameter, the
width of the Gaussian distribution, $\sigma$. This parameter is readily determined by
fitting the generated distribution to the one found at various experimental energies.
We consider specifically the pion rapidity distributions at SPS and RHIC energies.

\begin{figure}
\begin{center}
\includegraphics[scale=0.45]{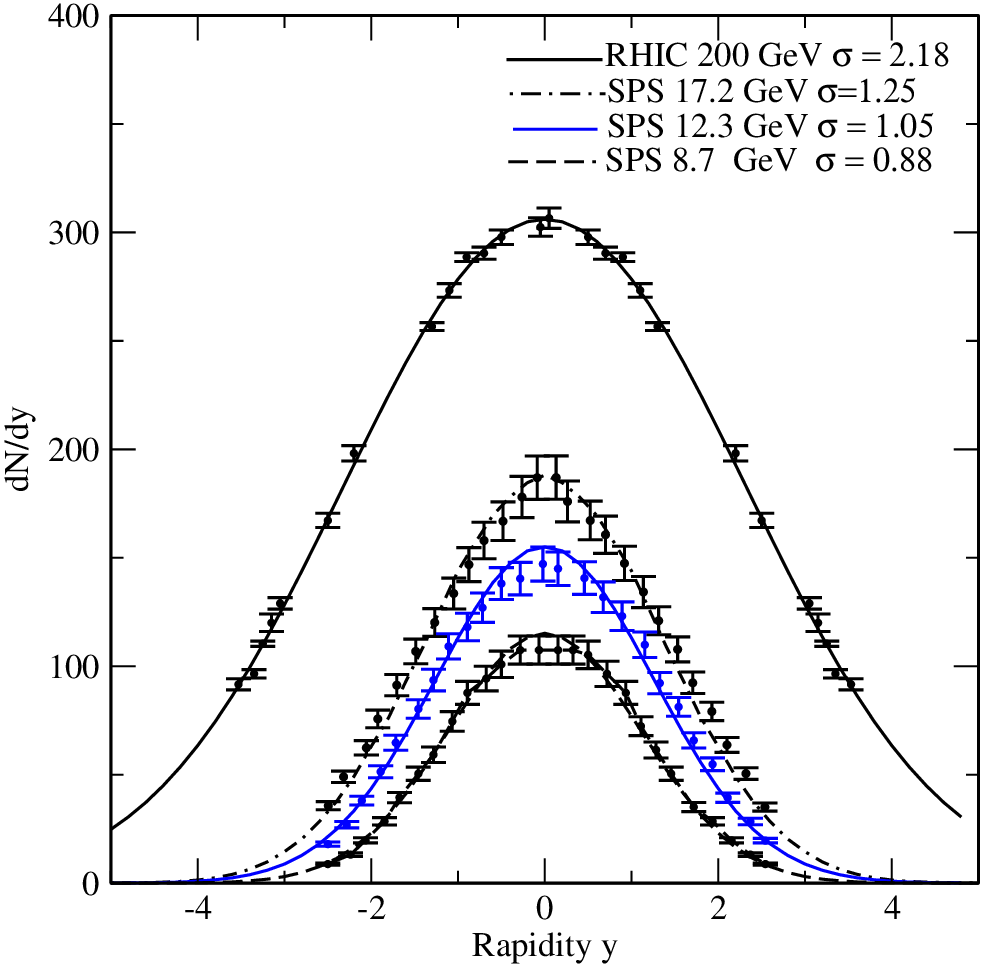}
\caption{(Color online) The pion rapidity spectra used to fit the Gaussian fireball distribution
width $\sigma$ to the experimental data from SPS and BRAHMS are shown.
}
\label{fits}
\end{center}
\end{figure}

The most abundantly produced particles in nuclear
collisions are pions. The pion rapidity distribution should then display the same features as the
total charged particle rapidity distribution. Since the baryon chemical potential has only a minimal
influence on the rapidity distribution of pions, this distribution is also an excellent candidate to determine
the fireball width. The experimental pion distributions were used to find the best
fit to $\sigma$ and these were then used to calculate the rapidity spectra coming from the
extended thermal model.

\begin{figure}
\begin{center}
\includegraphics[scale=0.45]{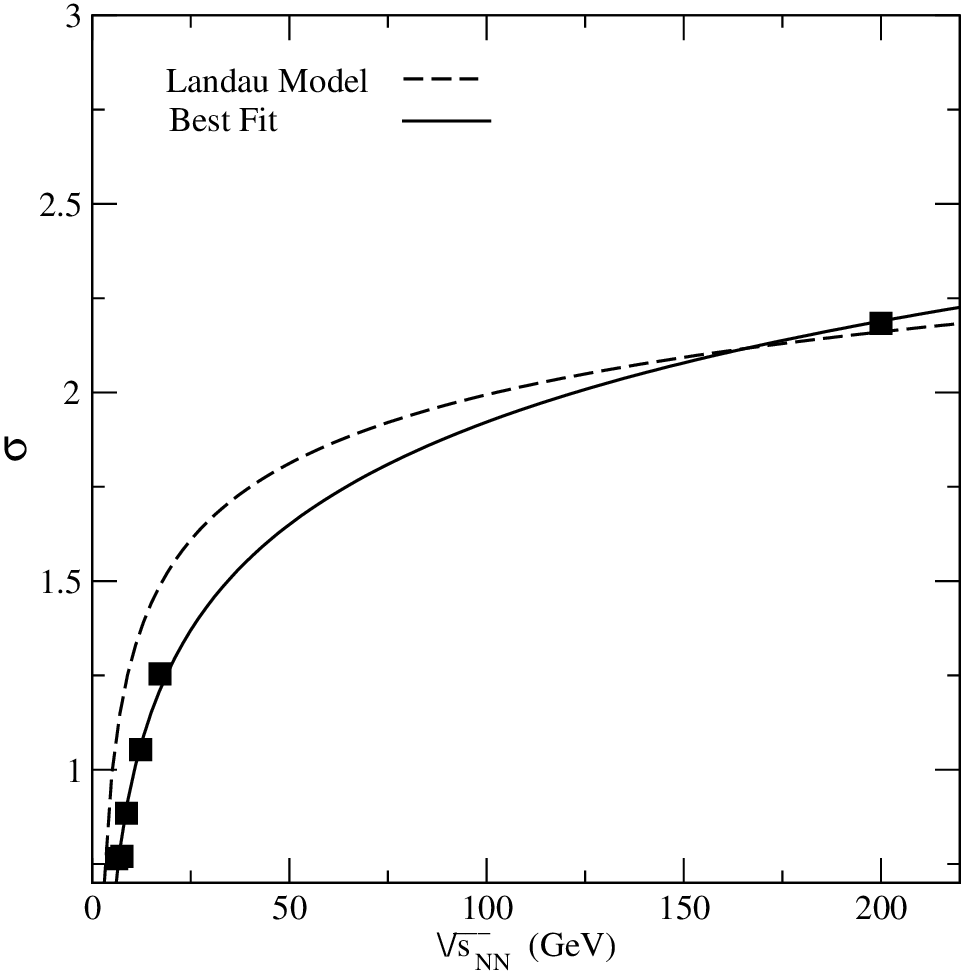}
\caption{Energy dependence of the distribution width.
The data points shown are the $\sigma$'s in the thermal model fitted to experimental data.
The short-dashed line is the prediction  based on the Landau model ($\sigma^2 \approx \ln \sqrt{s_{NN}}/2m_P$).}
\label{sigmaplt}
\end{center}
\end{figure}

The results of the fits for the rapidity spectra are shown in figure \ref{sigmaplt}.
The experimental distribution width has been proposed \cite{Bearden:2004yx} to change with
collision energy like $\sigma^2 = \ln \frac{\sqrt{s_{NN}}}{2 m_p}$  .
This is also shown in figure \ref{sigmaplt} and it can be seen that the two distribution
widths have a similar energy dependence. It is a remarkable property that this simple analytical
$\sqrt{s_{NN}}$ fit is applicable to heavy-ion collisions over such a wide range of beam energies.
It appeared for the first time in the classic  paper by Landau \cite{Landau} where the notion of
hydrodynamical evolution of a hadronic system was introduced . This concept was later
successfully used for the description of high energy multiparticle production in $pp$
collisions \cite{Carruthers:1973ws,Carruthers:1973rw,Cooper:1973vc}.

The differences seen in figure \ref{sigmaplt} could be attributed to specific heavy-ion processes
such as the cooling, freezing, and evaporation of the primary highly excited blob of dense
hadronic matter.

\begin{figure}
\begin{center}
\includegraphics[scale=0.45]{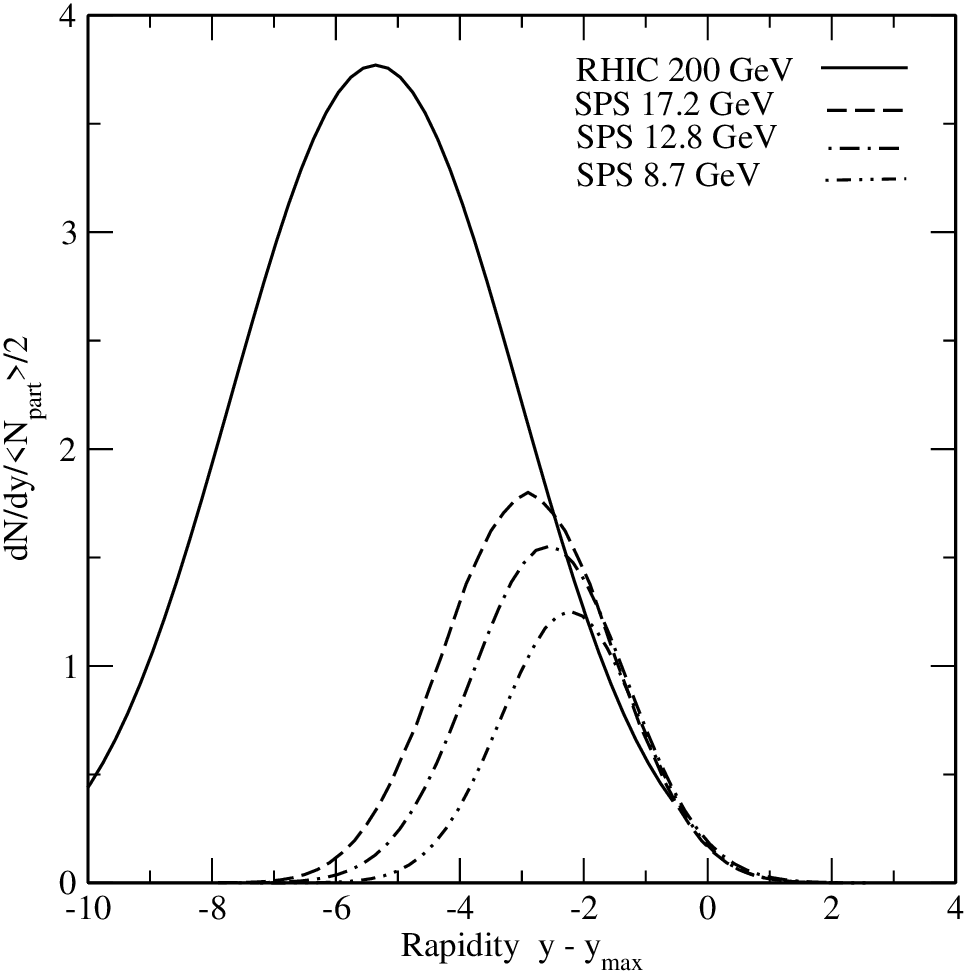}
\caption{Longitudinal scaling at the lower energies.
The values of $y_{\text{max}}$ are 5.36, 2.9, 2.569 and
2.222 at RHIC 200 GeV, SPS 17.2 GeV, SPS 12.3 GeV and SPS 8.7 GeV respectively.}
\label{low_scaling}
\end{center}
\end{figure}

For a comparison with LHC
predictions, $\frac{dN_{\pi}}{dy}/\frac{\langle N_{\text{part}}\rangle}{2}$ is plotted in
figure \ref{low_scaling}. Here we can clearly see the similar tails of the four rapidity
distributions for the higher SPS energies and for RHIC.
This is also seen in the experimental data in \cite{Busza:2004mc}.

\begin{figure}
\begin{center}
\includegraphics[scale=0.45]{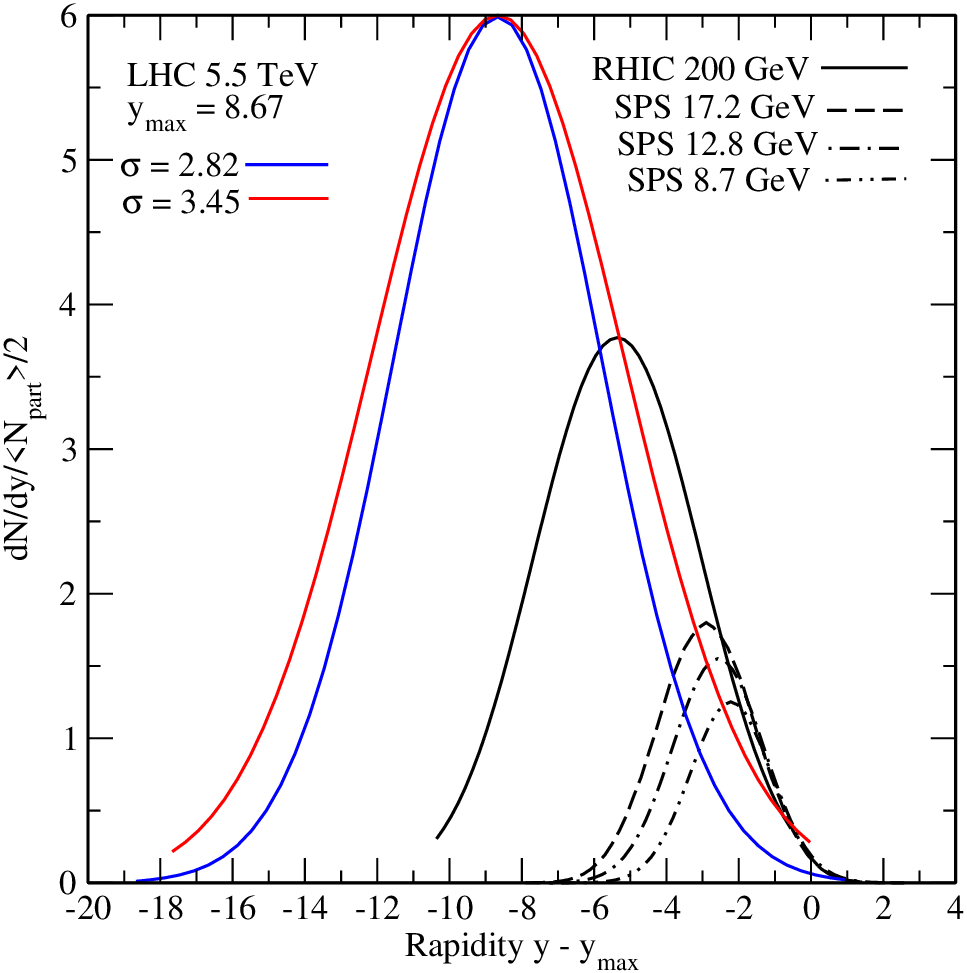}
\caption{(Color online) Extrapolating to LHC energies gives the highest curves in the figure
above ($\sigma = 3.45$ and $\sigma = 2.82$). The values of $y_{\text{max}}$ at RHIC and SPS
are the same as those given in Fig. 3.}
\label{hi_scaling}
\end{center}
\end{figure}

By extrapolating the results for $\sigma$ to LHC energies, we can make an extrapolation
based on the thermal model for the rapidity distribution.
Using the fitted curve we obtain $\sigma_{LHC} = 3.45$. Following the prediction based on
the Landau model curve one would obtain for LHC $\sigma = 2.82$. The extrapolation is over a
large energy range and thus both values of $\sigma$ are shown in
figure \ref{hi_scaling} for comparison.

It can clearly be seen that extended longitudinal scaling does not occur at LHC energies. Thus
the extension of the thermal model to describe rapidity distributions
is consistent with the concept of extended  longitudinal
scaling at SPS and RHIC energies but  not at LHC energies.

A violation of extended longitudinal scaling at LHC energies is also predicted in the
string percolation model \cite{Brogueira:2006nz}.

\section{Conclusions}
About thirty-five years ago it was noted in  \cite{Carruthers:1973ws} that the ``possible experimental fact
of a Gaussian rapidity distribution of produced particle is significant independently
of the Landau model. Only further detailed calculations of correlations and other
fine structure can be expected to establish or disprove the hydrodynamic picture of particle production.''

This statement was made  to describe the pending $p-\bar{p}$ experiments
at $\sqrt{s_{NN}}=53$ GeV  at that time. The outcome is
still pending concerning the $\sqrt{s_{NN}}= 5.5 - 14$ TeV present day
experiments.

\begin{acknowledgments}
This work was supported in part by the Scientific and Technological Co-operation
Programme between Poland and South Africa and in part by the Polish Ministry of
Science and  Higher Education under contract No.~N~N202~0953~33.
\end{acknowledgments}

\end{document}